\documentclass{aa}  

\usepackage{graphicx}
\usepackage{txfonts}
\usepackage{xcolor,colortbl}
\usepackage{hyperref}
\usepackage{multirow}
\usepackage[normalem]{ulem}
\usepackage{txfonts}
\usepackage{upgreek}
\DeclareUnicodeCharacter{2212}{\ensuremath{-}}
\makeatletter
\renewcommand*\aa@pageof{, page \thepage{} of \pageref*{LastPage}}
\makeatother

\begin{document}

   \title{The Clouds in the "Middle Ages"}
   \subtitle{I. Star formation history and chemical abundances in the LMC}

   \author{M. Schultheis
            \inst{1,2}
            \and
            M.-R. L. Cioni
            \inst{2}
            \and
            S. Anoardo
            \inst{2,3}
    }
   \institute{Universit\'{e} de la C\^{o}te d’Azur, Observatoire de la C\^{o}te d’Azur, Laboratoire Lagrange, CNRS, Blvd de l’Observatoire, 06304 Nice, France
    \email{mathias.schultheis@oca.eu}
    \and
    Leibniz-Institut für Astrophysik Potsdam, An der Sternwarte 16, D-14482 Potsdam, Germany 
    \and
    Institut f\"{u}r Physik und Astronomie, Universit\"{a}t Potsdam, Haus 28, Karl-Liebknecht-Str. 24/25, D-14476 Potsdam, Germany 
    }

   \date{Received July 24, 2026; accepted Month Day, Year}

\abstract{The Large Magellanic Cloud (LMC) is a benchmark galaxy for studying stellar evolution, chemical enrichment, and galaxy interactions due to its proximity and intermediate metallicity. While its star formation history (SFH) has been extensively studied using colour-magnitude diagram (CMD) fitting techniques and the chemical enrichment is largely based on red giant branch (RGB) stars, the asymptotic giant branch (AGB) population—a key tracer of intermediate-ages—has received less attention in these analyses.}
{We aim to reconstruct the SFH of the LMC using AGB stars and to investigate their reliability as tracers of chemical abundances.}
{The initial sample of 879 AGB stars from the Gaia DR3 long-period variable catalog was extended to 2047 fundamental-mode pulsators by applying separate O-rich and C-rich period-Wesenheit relations; OGLE-III cross-matching then showed that only 256 of these stars are genuine Mira variables. We reconstructed the SFH using two independent period-age calibrations: the theoretical relation of \citet{Trabucchi2022}  and the empirical calibration of \citet{Zhang2023}. For the chemical abundance analysis, we used APOGEE DR17 spectra for 296 AGB stars and 271 RGB stars, applying a refined line-by-line abundance pipeline with a tailored line selection to derive Mg, Si, and Ca abundances.}
{The two period-age calibrations yield significantly different SFHs, and neither matches independent CMD-based results equally well, so we present both rather than one fiducial choice. Our SFH reveals three distinct phases: an ancient population at 9−10 Gyr, a quiescent phase at 4–8 Gyr, and a starburst peaking at 1.5−2 Gyr. Our chemical analysis demonstrates that carefully selected Mg, Si, and Ca lines yield significantly reduced intrinsic scatter compared to APOGEE DR17 results.}
{AGB stars prove to be reliable tracers of both the SFH and chemical enrichment of the LMC, provided that careful line selection and calibration are applied.}


   \maketitle

\section{Introduction}

The formation and evolution of galaxies is driven by two processes, the progress of star formation from the cold gas and the accretion resulting from the interaction with other systems. The Large and Small Magellanic Clouds (LMC and SMC) are the most massive and brightest dwarf galaxies and probably approached the Milky Way only a few Gyr ago (e.g. \citealt{Patel2017}, but see \citealt{Vasiliev2024} for a second passage scenario). The Clouds have also interacted with each other multiple times (e.g. \citealt{Besla2012},  \citealt{Diaz2012}, \citealt{Hammer2015},  \citealt{Cullinane2022}) and due to their proximity (50--60 kpc; \citealt{deGrijs2014}, \citealt{deGrijs2015}) they represent a laboratory to study the consequences of dwarf-dwarf galaxies interactions in detail, and the interaction with a massive host. 
Due to their relatively low average metallicity (about one-half and one-fifth of the solar iron abundance for the LMC and SMC, respectively) they are also fundamental benchmarks for stellar properties and stellar evolution. 

The most intense period of star formation in the LMC occurred 0.5--4 Gyr ago (e.g. \citealt{Ruiz-Lara2020}, \citealt{Mazzi2021}, \citealt{Burhenne2026}) and a steep increase in the number of star clusters confirmed a burst 2--3 Gyr ago (e.g. \citealt{Pieres2016}). During this time, distinct peaks of star formation also occurred in the SMC (\citealt{Massana2022}) and its star clusters (\citealt{Dhanush2024}), suggesting a link between enhanced star formation and tidal interactions, whereas at ages $\rm > 5\,Gyr$ the LMC and SMC might have evolved independently. Most of the SMC mass was assembled prior to 3.5 Gyr ago \citep{rubele2018}.  Morphological maps of high spatial resolution (about 0.15 kpc) show a well-defined LMC bar with a clear central density embedded in a disc-like distribution, a major SMC inner over-density and several protuberances \citep{elyoussoufi2019}, depending on the stellar population. The periphery of the Clouds reveals also a wealth of substructures in kinematics and metallicity which were mostly produced by multiple LMC-SMC interactions (e.g., \citealp{belokurov2019}). The formation of northern and southern LMC arms are likely associated to an interaction at about 1 Gyr ago whereas the western LMC truncation and eastern SMC extensions could be associated to interactions at about 400 and 150 Myr ago, respectively (e.g. \citealp{Cullinane2022}). Metallicity maps confirmed a metallicity gradient in the LMC (e.g. \citealp{Grady2021}, \citealp{omkumar2025}, \citealp{Povick2025b,Povick2025a}) and various asymmetries, including a metal-rich ring-like feature possibly of intermediate age \citep{choi2018}. The stellar motion within morphological and metallicity features depicts the disruption of the SMC as imprinted both on its outer and inner regions (e.g.  \citealp{Vijayasree2026}).

Asymptotic giant branch (AGB) stars, are typical intermediate-age tracers of the Clouds, as the majority of the red giant branch (RGB) and red clump stars in the Clouds are on average >3 Gyr old, see \cite{elyoussoufi2019} for the median ages of the different stellar populations. AGB stars are low- to intermediate-mass stars experiencing shell He and H burning with an atmosphere enriched in carbon and/or oxygen products and a large envelope where dust may also form. Atomic and molecular features dominate the spectra of carbon- (or C-rich) and oxygen- (or O-rich) AGB stars and unmistakably allow to distinguish between the two types. The formation of carbon stars is directly linked to the number and efficiency of the third dredge up process which brings elements synthesised near the core to the surface. This process is more efficient at low metallicity (see e.g. \citealp{karakas2002}) and in stars with masses between $\sim$1.5 and 3 M$_\odot$ \citep{pastorelli2020}, which means that carbon stars trace a specific age interval (from about 300 Myr to 2 Gyr). On the contrary, O-rich AGB stars span a larger age range (up to 10 Gyr). Most AGB stars intrinsically pulsate with long periods (from a few tens to several hundred days) and large amplitudes (from a fraction to a few magnitudes depending on waveband). Their pulsations are powered by convective processes (see e.g. \citealt{Freytag2017}). Furthermore, they are notable producers of dust and contribute, due to their strong stellar winds, to both the chemical enrichment of galaxies and the integrated luminosity of intermediate-age stellar populations (e.g., \citealp{karakas2014,matsuura2009,maraston2006}). These properties corroborate their importance as tracers of stellar evolution, stellar populations and distances (e.g. \citealp{pastorelli2020,madore2020,iwanek2021,Trabucchi2022}).

Among AGB pulsators, Mira variables offer a valuable means of determining ages (see e.g. \citealt{Sanders2024}) . These stars experience thermal pulsations in the fundamental mode, with periods ranging from 80 to 1000 days and typical K-band amplitudes of at least 0.4 mag (\citealt{Habing2004}).  Miras are known to exhibit well-defined period–luminosity (PL) relationships (see e.g. \citealt{Glass1981}; \citealt{Feast1989}; \citealt{Ita2004}, \citealt{Ita2011}), because stars of a fixed mass only enter the fundamental-mode pulsation stage within a restricted range of radii (\citealt{trabucchi2019}). For analogous reasons, theory predicts that Miras should obey a period–age relation (see e.g. \citealt{Feast1987}, \citealt{Trabucchi2022}). \citet{Trabucchi2022} combined nonlinear pulsation calculations with isochrone models. These models predict that the fundamental mode period decreases with increasing age and the authors give for O-rich and C-rich models the corresponding best-fit relations.

Observational support for such a relation has existed for decades, stemming from observed links between Mira periods and their kinematic properties, as well as between periods and population scale heights (\citealt{Feast1963}). Since hotter kinematics generally correspond to older stellar populations, these correlations are understood as evidence of a period–age connection, further validated by studies of Mira variables in stellar clusters (\citealt{Grady2019},  \citealt{Zhang2023}). Empirical calibrations have been derived from data in the solar neighbourhood  (\citealt{Feast2000},   \citealt{Catchpole2016}, \citealt{Grady2020}). More recently, \citet{Zhang2023} fitted dynamical models to Gaia DR3 Mira variables, linking their radial and vertical velocity dispersions to their period, and then comparing these dispersions with age-dependent dispersions from main-sequence turn-off and subgiant stars. This allowed them to derive a period–age relation consistent with earlier work and with cluster member data. Based on this work, \citet{Sanders2024} traced the SFH of Mira variables in the nuclear stellar disc (see also \citealt{Schultheis2025}). 

Determining SFHs from long-period variables (LPVs) has the advantage that sources are relatively bright, but are less numerous, than for SFH studies based on fitting the colour-magnitude-diagram (CMD) resulting from deep imaging (e.g. \citealt{Ruiz-Lara2020}, \citealt{Mazzi2021}). However, a time-domain survey is needed to obtain well-sampled light-curves for determining accurate periods -- especially long ones -- whereas for CMD-fitting single-epoch two-waveband observations do usually suffice. Previously, \citet{Rezaeikh2014} derived the SFH from the K-band magnitude distribution of LPVs,  which was converted to a mass-age relation using the stellar models from \citet{Marigo2008}. In that case, sparsely sampled light-curves were used to identify the LPV nature of the sources. In this project, we will take advantage of period-age relations to trace the SFH in the LMC by applying a similar method as in \citet{Sanders2024}, since the calibrated period-age relations demonstrate that Mira variables span ages from 1--10 Gyr making them excellent probes of the formation and distribution of the stellar population. The two approaches also differ in their systematic sensitivities. The luminosity-function method of \citet{Rezaeikh2014} requires an assumed initial mass function (IMF) to invert the observed luminosity function into a star-formation rate, and relies on a bolometric luminosity-age relation that is itself strongly metallicity dependent; it further requires deep, highly complete photometry, since the oldest, faintest AGB stars are the most readily lost to incompleteness. The period-age approach instead assigns each star an age directly from its pulsation period: IMF dependence enters only through the exponent weighting the period-dependent duration of the Mira phase (Sect.~\ref{sect:SFH}), which we show has a negligible effect on our results.

This is the first work from the Clouds in the "Middle Ages" project that aims to obtain line-sight-velocities, stellar parameters and elemental abundances of AGB stars in the Clouds for tracing  kinematical and chemical patterns associated to their SFHs. The patterns of elemental abundances and their gradients, with respect to similar quantities derived from RGB stars and younger stars, will produce vital information to develop chemical evolution models and explain the nature of structural features across the galaxies, supported or contrasted by the dynamical behaviour of the stars derived from the line-of-sight velocities and the proper motions. Galactic chemical evolution tries to explain how the chemical elements have formed and their abundances have evolved in time and distributed in space. In particular, they aim at following the evolution of the gas and its chemical composition inside galaxies. Such an evolution depends on the SFH, stellar nucleosynthesis and possible gas flows. 

The paper is organised as follows. Sect.\,\ref{lpv-sfh} shows the calculation of stellar ages for a subset of AGB stars, Mira variables, and the resulting SFH. Section \ref{abundances} presents the computation of Mg, Si and Ca abundances from APOGEE spectra. Finally, Sect.\,\ref{conclusions} provides a perspective on upcoming analyses  of the Clouds.

\section{Mira variables as indicators of stellar ages} \label{lpv-sfh}

\subsection{Data} \label{data}

We use the  Gaia DR3 Long-Period Variable (LPV) star catalog of \citet{Lebzelter2023}, which is the largest all-sky LPV catalog to date. Based on 34 months of Gaia data, this catalog contains about 1.7 million LPVs over the full sky for a  period range between 35 -- 1\,000 days and amplitudes in the G band larger than 0.1\,mag. Based on the Red Photometer (RP; 640--1050 nm and R$~$50) spectra, they performed a classification of C stars which are flagged in the Gaia LPV catalog.

We have selected our LMC sample as in \citet{Nidever2020}, using equatorial coordinates and a radial velocity selection ($\rm 160 < v_{rad} < 350$\,km/s) to ensure membership to the LMC. Our sample consists of 9588 LPVs which we cross-matched with the 2MASS catalogue. We include C-rich and O-rich stars using the Gaia DR3 classification flag $is\_cstar$.

LPVs occupy different locations in the PL relation based on their pulsation mechanism. \citet{Wood2000} demonstrated that the different PL sequences correspond to different radial pulsation modes. Miras were unambiguously identified as fundamental mode pulsators on sequence C, while semiregular variables on sequences A and B were found to pulsate in the higher overtones. Subsequent refinements introduced additional sequences (A$^\prime$, C$^\prime$) and suggested that sequences such as A and B are composed of multiple, closely-spaced sub-sequences (e.g. \citealt{Ita2004}).
\citet{Soszynski2007} found a more precise characterisation of the PL relation based on the Wesenheit index ($\rm W_{K}$) which is a reddening independent quantity and therefore the period-$\rm W_{K}$ relations can be used as distance indicators even for heavily reddened Mira variables.

\begin{figure}[!htbp]
 \center{\includegraphics[width=0.45\textwidth]{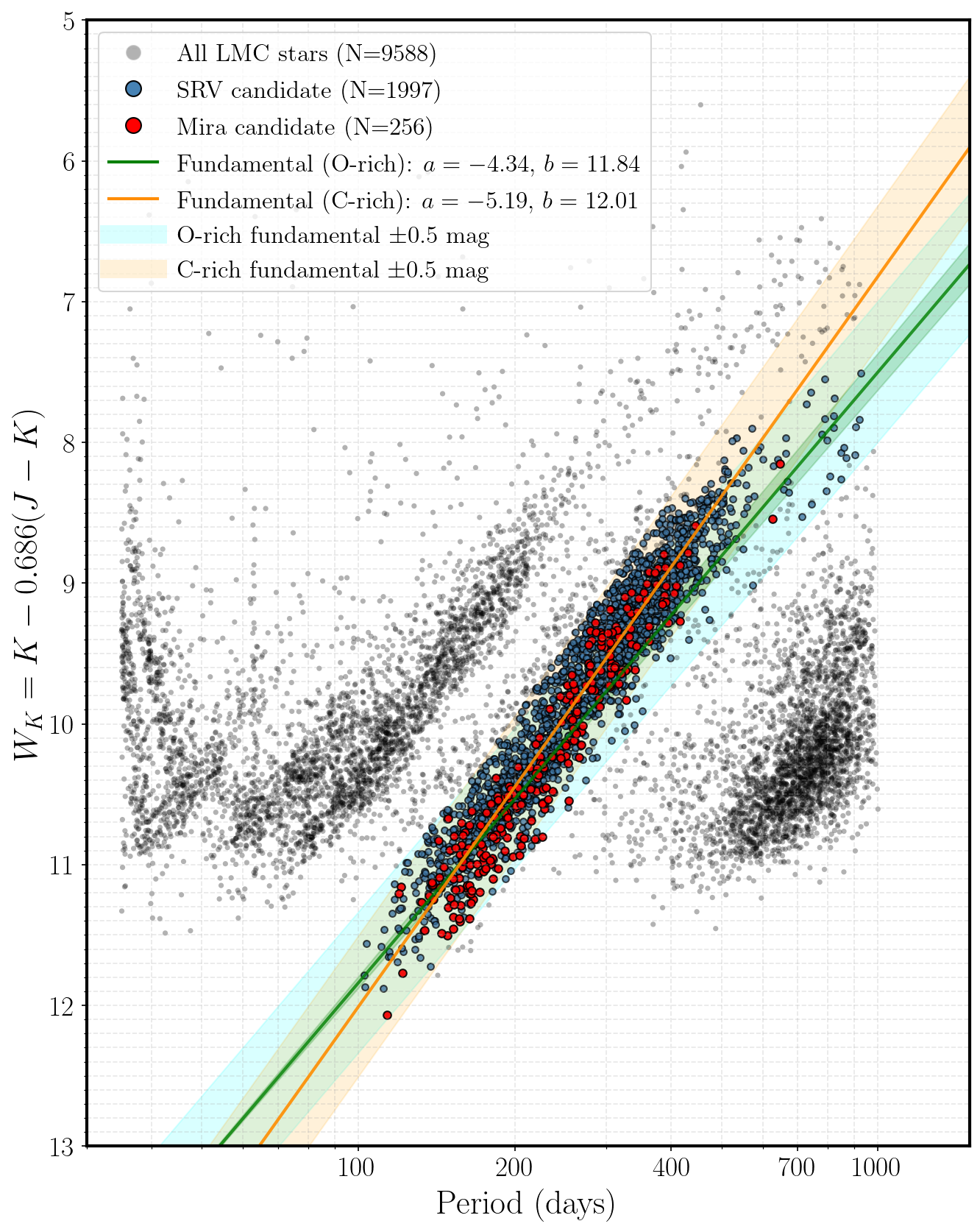} }
 \caption{LMC sample of the LPV Gaia DR3 catalog (\citealt{Lebzelter2023}). The y-axis denotes the Wesenheit function as $\rm W_{K} = K - 0.686*(J-K)$ where J and K are the 2MASS magnitudes, while the x-axis shows the period (days). Red circles indicate Mira variables while blue circles indicate SRVs. The green straight line shows the fundamental mode relation from \citet{Soszynski2007} for O-rich stars with the $\rm 1\,\sigma$ uncertainties while the orange line is for C-rich stars. The cyan and orange areas denote our selection criteria of fundamental mode pulsators with $\rm \pm\,0.5\,mag$ around the fundamental mode relation (for O-rich and C-rich stars).}
 \label{Wesenheit}
 \end{figure}

Figure~\ref{Wesenheit} shows the period-$\rm W_{K}$ diagram of our LMC sample  revealing the different pulsational sequences, and in  particular the fundamental mode sequences (indicated by the  green line for O-rich  and the orange line for C-rich LPVs, \citealt{Soszynski2007} -- with their uncertainties). We have selected our sample of fundamental pulsating LPVs within $\rm \pm 0.5\,mag$ around these relations (indicated as the filled areas in Fig.~\ref{Wesenheit}). Note that these sequences also include fundamental-mode pulsating semiregular variables (SRVs). This sample consists of 2047 stars with  a mean period of $\rm \sim 290\,days$ and a mean peak-to-peak amplitude in G of  $\rm \sim 0.6\,mag$.

As we will use the period-age relations for Mira variables to retrieve the SFH (see Sect.~\ref{sect:SFH}),  we cross-matched this sample against the OGLE-III LMC catalogue of LPVs (\citealt{soszynski2009}), which classifies sequence-C pulsators into genuine Miras and SRVs directly from well-sampled I-band light curves, using the standard peak-to-peak amplitude boundary $\rm \Delta I > 0.8\,mag$. We derived an empirical relation: $\Delta G = 0.149 + 0.959\times\Delta I$  with a residual scatter of 0.18 mag. Applying this relation to OGLE's own $\Delta I > 0.8$ mag Mira boundary gives an equivalent boundary of $\Delta G \gtrsim 0.91$ mag for our sample (see Sect.\,\ref{amplitude} for further details). We find that only 251 stars ($\rm \sim 12\%$) meet this Mira criterion.

\subsection{Star formation history} \label{sect:SFH}

We make use of the period-age relation for Mira variables to construct the SFH of the LMC following the approach of \citet{Sanders2024}. We employ two distinct period-age calibrations. The first is the polynomial relation of \citet{Trabucchi2022}, based on theoretical stellar evolution models:

\begin{equation}
\log_{10}(\tau) = 10.78 - 2.660\left(\frac{P}{350.0}\right) + 0.5953\left(\frac{P}{350.0}\right)^2
\label{eq:trabucchi}
\end{equation}

where $\tau$ is the stellar age in years and $P$ is the pulsation period in days. The second is the empirical sigmoidal calibration of \citet{Zhang2023}:

\begin{equation}
\tau = 7.35 \left[1 + \tanh\left(\frac{330 - P}{308}\right)\right]
\label{eq:zhang}
\end{equation}

where $\tau$ is in Gyr. This relation was calibrated exclusively on O-rich, kinematically-selected Miras in the Milky Way disc; the \citet{Zhang2023} sample explicitly excludes SRVs via an amplitude and period selection as well as C-rich stars. We therefore evaluate this calibration only on the subset of our sample satisfying \citet{Zhang2023}'s own published O-rich amplitude and period selection criteria, whereas the \citet{Trabucchi2022} relation, whose calibrating sample included both Miras and SRVs as well as O- and C-rich stars, is applied to the full chemistry-inclusive sample (Sect.~\ref{data}).

It is worth noting that, independently of the amplitude selection, the two calibrations also differ in the population and host galaxy they were calibrated on. \citet{Trabucchi2022} calibrate their relation on $\sim$95 fundamental-mode pulsators in star clusters spanning the Galaxy, the LMC (19 clusters), and the SMC (NGC 419), with ages taken from published cluster isochrone fits; the authors explicitly state the relation is "environment-dependent, and not necessarily universal". \citet{Zhang2023}, by contrast, calibrate their relation from the kinematics of $\sim$8300 field Mira candidates in the Milky Way disc, using a dynamical model to convert velocity dispersion in period bins into ages; their own spatial selection explicitly excludes the LMC's sky position and distance range from that sample. Their relation was therefore calibrated on zero LMC stars, using Milky Way disc dynamics, and applying it to the LMC assumes the period-age relation is portable across host galaxies and environments — an assumption independent of, and larger than, the amplitude selection.

To properly propagate observational and systematic uncertainties, we implement a Monte Carlo (MC) simulation with $N_{\mathrm{MC}} = 100$ iterations. Each iteration applies three independent perturbations to the data.

First, period measurement uncertainties are propagated through Gaussian perturbations:

\begin{equation}
P_{i}^{\mathrm{pert}} \sim \mathcal{N}(P_i, \sigma_{P_i}^2)
\label{eq:period_pert}
\end{equation}

where $\sigma_{P_i}$ is the measured period uncertainty for star $i$. This accounts for observational errors in the period determination.

Second, completeness corrections account for the sample's detection efficiency. We assume a typical 90\% completeness based on the recovery rate of Gaia LPVs with respect to OGLE and ASAS-SN data (\citealt{Lebzelter2023}), modeled as:

\begin{equation}
C \sim \mathcal{N}(0.90, 0.09)
\label{eq:completeness}
\end{equation}

clipped to the range $[0.70, 1.00]$ to avoid unphysical values.

Third, bootstrap resampling accounts for statistical uncertainty arising from the finite sample size. For each MC iteration, we generate a new dataset of equal size to the original by randomly drawing stellar indices with replacement from the parent sample. This procedure naturally propagates Poisson noise through the entire SFH reconstruction pipeline—including period-age conversion, weighting, and binning—without making assumptions about the underlying age distribution.

To convert the number density in age into a star formation rate, we follow \citet{Sanders2024}, who showed that the number of Mira variables per unit age is proportional to $\tau^{0.48}$ times the star formation rate (SFR):

\begin{equation}
\frac{\mathrm{d}N}{\mathrm{d}\tau} \propto \tau^{0.48} \times \mathrm{SFR}(\tau)
\label{eq:sfr_relation}
\end{equation}

\begin{figure}
    \centering
    \includegraphics[width=0.49\textwidth]{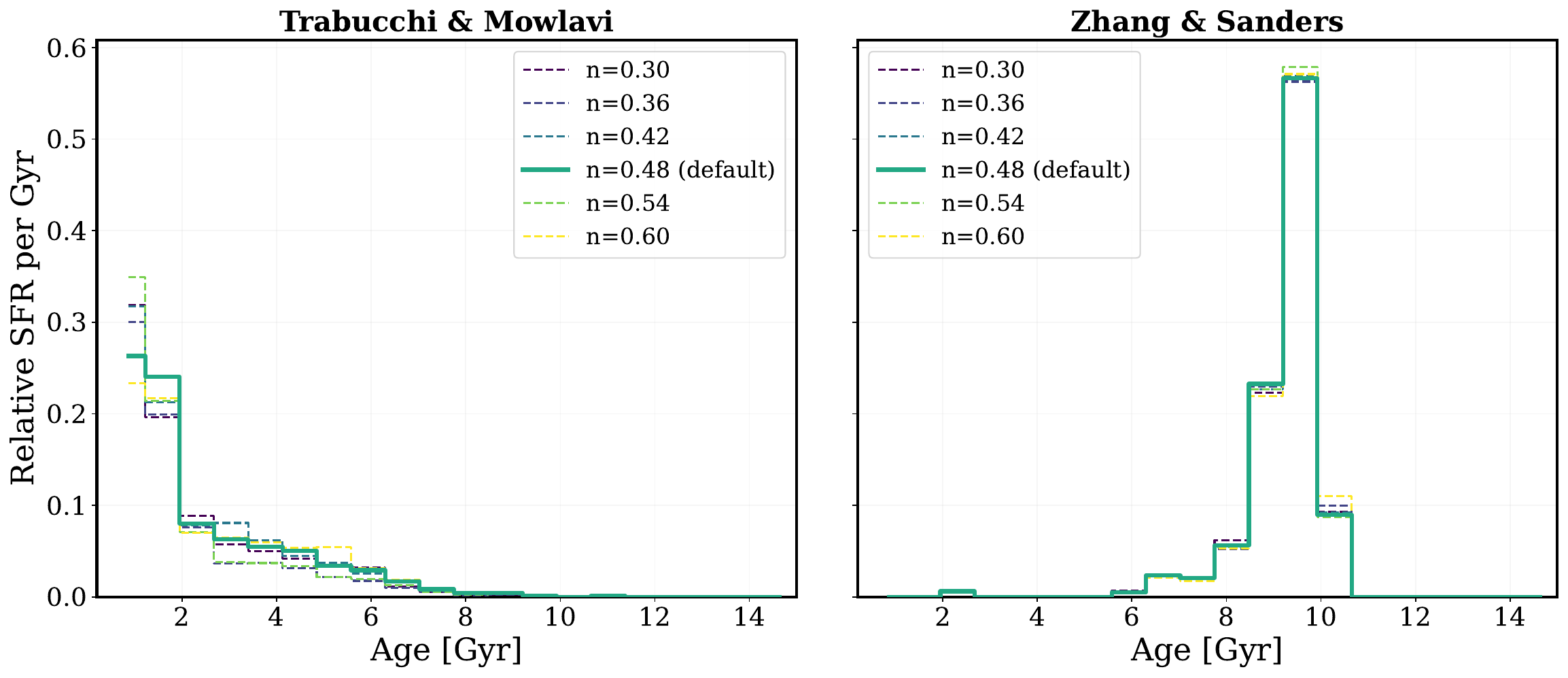}
    \caption{SFH recomputed at six values of the $\tau^n$ exponent spanning $n=0.30$–$0.60$; the default value corresponds to \cite{Sanders2024} for the two calibrations used in this study.}
    \label{fig:exponent}
\end{figure}

The $\tau^{0.48}$ factor arises from the combination of the initial mass function (IMF) and the mass-dependent duration of the Mira phase, which we assume to be  constant with age. Because both the IMF slope and the mass-dependent Mira-phase duration are metallicity dependent, and this exponent was derived by \citet{Sanders2024} for the Milky Way's nuclear stellar disc -- a different environment, star-formation regime, and metallicity from the LMC -- we tested its impact by recomputing the SFH for $n \in [0.30, 0.60]$. The resulting star-formation-rate-weighted mean age changes by less than 10\% for the \citet{Trabucchi2022} calibration and by less than 1\% for the \citet{Zhang2023} calibration over this range, and none of our qualitative conclusions depend on the specific value adopted (Fig.\,\ref{fig:exponent}).

For each MC iteration, the total weight for each star is:

\begin{equation}
w_i = \tau_i^{0.48} \times C^{-1} \times \exp\left[s(\log P_i)\right]
\label{eq:weight}
\end{equation}

where $C$ is the survey completeness factor, and $s(\log P)$ is a spline-based systematic correction accounting for uncertainties in the period-age relation. The spline correction is modeled as a Gaussian process with covariance:

\begin{equation}
\begin{split}
\mathrm{Cov}\left[s(\log P), s(\log P')\right] = &\; \sigma_{\mathrm{spline}}^2 \, \delta(\log P - \log P') \\
& + \sigma_{\mathrm{corr}}^2 \exp\left(-\frac{|\log P - \log P'|}{\ell}\right)
\end{split}
\label{eq:spline_cov}
\end{equation}

with parameters $\sigma_{\mathrm{spline}} = 0.03$, $\sigma_{\mathrm{corr}} = 0.005$, and correlation length $\ell = 2$ in knot space. This captures both independent and correlated systematic uncertainties.

The total SFH in each age bin $j$ is then computed as:

\begin{equation}
\mathrm{SFR}_j = \frac{\sum_i w_i \, \mathbb{I}(\tau_i \in \mathrm{bin}_j)}{\sum_i w_i}
\label{eq:sfh}
\end{equation}

where $\mathbb{I}(\tau_i \in \mathrm{bin}_j)$ is the indicator function that equals unity if the age $\tau_i$ falls within the age interval defining the $j$-th bin, and zero otherwise. The denominator ensures normalization where

\begin{equation}
\sum_j \mathrm{SFR}_j = 1
\label{eq:normalization}
\end{equation}

represents the relative SFR per Gyr.

Uncertainties are propagated via the MC simulations, with final uncertainties given by the 16th and 84th percentiles of the resulting distributions:

\begin{align}
\mathrm{SFR}_j &= \mathrm{median}\left(\left\{\mathrm{SFR}_j^{(k)}\right\}_{k=1}^{N_{\mathrm{MC}}}\right) \\
\sigma_j^{\mathrm{lower}} &= \mathrm{percentile}\left(\left\{\mathrm{SFR}_j^{(k)}\right\}_{k=1}^{N_{\mathrm{MC}}}, 16\right) \\
\sigma_j^{\mathrm{upper}} &= \mathrm{percentile}\left(\left\{\mathrm{SFR}_j^{(k)}\right\}_{k=1}^{N_{\mathrm{MC}}}, 84\right)
\label{eq:uncertainties}
\end{align}

\noindent where $\mathrm{SFR}_j^{(k)}$ is the SFH in bin $j$ computed from the $k$-th MC iteration.

\begin{figure}[!htbp]
 \center{\includegraphics[width=0.45\textwidth]{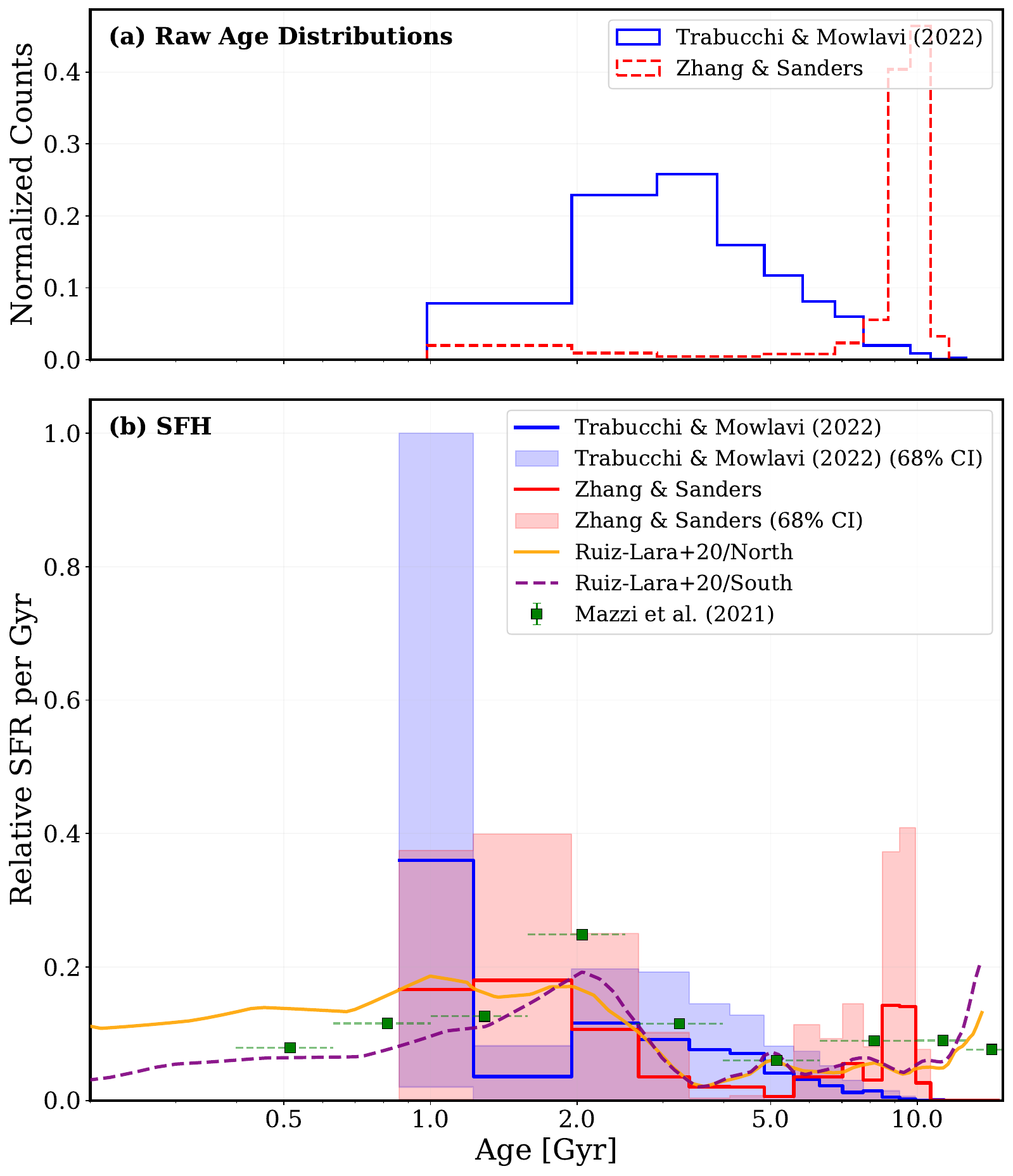} }
 \caption{Reconstructed SFH of the LMC derived from the full fundamental-mode-pulsator sample (Sect.~\ref{data}), i.e. without applying the amplitude criterion used to isolate genuine Mira variables. The blue solid line and shaded region show the \citet{Trabucchi2022} calibration with 68\% confidence intervals; the red dashed line and shaded region show the \citet{Zhang2023} calibration. Green points with error bars show external data based on CMD-fitting techniques from \citet{Mazzi2021}; orange solid and purple dashed lines show the \citet{Ruiz-Lara2020} North and South regions, respectively. The x-axis shows the age in Gyr;  The SFH is normalized such that the integral over all bins equals unity.}
 \label{fig:sfh_all}\label{fig:sfh_all}
 \end{figure}

 \begin{figure}[!htbp]
 \center{\includegraphics[width=0.45\textwidth]{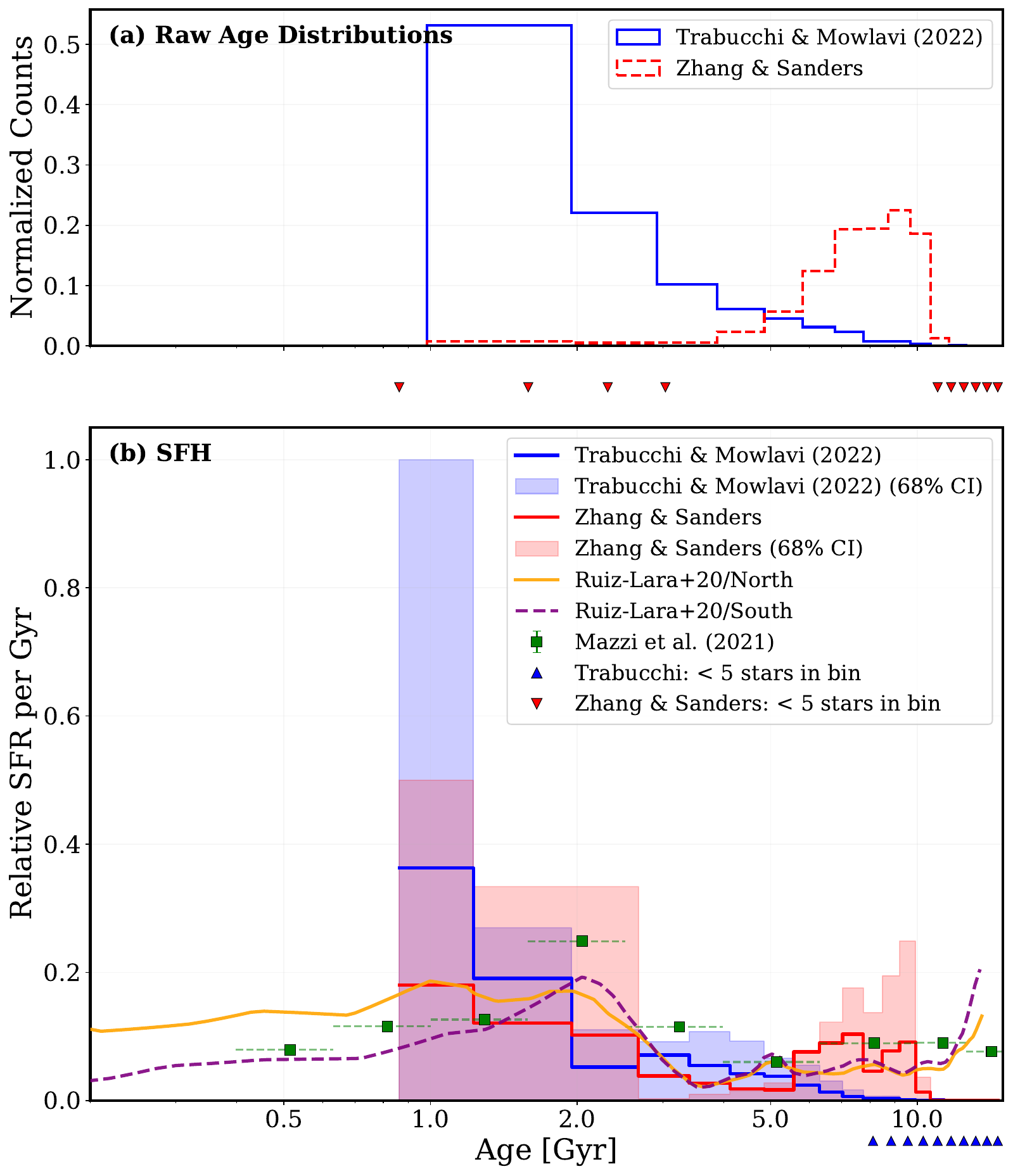} }
 \caption{As Fig.~\ref{fig:sfh_all}, but derived from the amplitude-selected Mira-only sample (Sect.~\ref{data}). Downward red triangles above the axis mark age bins containing fewer than five stars under the \citet{Zhang2023} calibration; upward blue triangles below the axis mark such bins under the \citet{Trabucchi2022} calibration.}
 \label{fig:sfh_mira}
 \end{figure}

\subsection{Results}  

Figures~\ref{fig:sfh_all} and~\ref{fig:sfh_mira} present the reconstructed SFH of the LMC, derived respectively from the full fundamental-mode-pulsator sample and from the amplitude-selected Mira-only sample (Sect.~\ref{data}), comparing the two period-age calibrations with independent SFHs.  \citet{Ruiz-Lara2020} used photometric data from the optical Survey of the Magellanic Stellar History (SMASH; \citealp{Nidever2021})  reaching sources down the  main sequence turnoff and adopting a CMD-fitting technique to derive the global SFH in the LMC. They found a distinct population in the spiral arms with an age of about 2\,Gyr. \citet{Mazzi2021} used the photometric near-infrared data from the VISTA Survey of the Magellanic Clouds system (VMC; \citealp{cioni2011}) and compared the observed CMDs with synthetic ones derived from stellar population analysis with TRILEGAL\footnote{https://stev.oapd.inaf.it/cgi-bin/trilegal}. They found that the period of most intense star formation occurred between 0.5 and 4\,Gyr, with an enhanced SFR at $\rm \sim1\,Gyr$.

A striking result of our analysis is the significant disagreement between the two period-age calibrations, with the \citet{Trabucchi2022} and \citet{Zhang2023} relations yielding substantially different SFH shapes, amplitudes, and peak ages. We find especially significant differences at  the oldest ages where \citet{Zhang2023} finds a peak in the SFR in the age-range of 8--10\,Gyr.

Neither calibration reproduces the external CMD-fitting SFHs of \citet{Mazzi2021} and \citet{Ruiz-Lara2020} equally well across all ages: the \citet{Trabucchi2022} relation more closely matches their intermediate-age (0.5--4 Gyr) SFR level and shape, while the \citet{Zhang2023} relation is more consistent with the substantial old population implied by the short, ancient formation episode found by \citet{Harris2009}. We therefore do not adopt either relation as a single fiducial calibration; instead we present results under both throughout, and treat their difference as our primary estimate of the systematic uncertainty associated with the choice of period-age calibration.

As a further check, we compared the SFH derived from our amplitude-selected Mira-only sample (Fig.~\ref{fig:sfh_mira}) with that obtained from the full fundamental-mode-pulsator sample (Fig.~\ref{fig:sfh_all}), i.e. without applying the amplitude criterion used to isolate genuine Miras from other fundamental-mode variables. The two samples agree on the qualitative pattern of disagreement between calibrations: under \citet{Trabucchi2022} both show their dominant contribution near 1 Gyr, and under \citet{Zhang2023} both show enhanced star formation in the oldest age bins. The amplitude selection concentrates these features and the two calibrations show a better agreement for ages <2 Gyr.

We note that, for the amplitude-selected Mira sample, the old age bins contain fewer than five stars each under both calibrations (marked explicitly in Fig.~\ref{fig:sfh_mira}); while the old-age excess is qualitatively recovered independently in both the Mira-only and the full fundamental-mode samples, its precise amplitude in the Mira-only sample should be interpreted with this small-number caveat in mind. We nonetheless adopt the amplitude-selected Mira sample as our primary result, since the period-age calibrations used here (\citealp{Trabucchi2022,Zhang2023}) are most directly applicable to fundamental-mode Mira pulsators specifically.

The Mira-based SFH reveals three distinct phases of star formation activity, which we discuss in detail below. The oldest component of the LMC's SFH contributes approximately 20\% of the total SFR formed over the LMC's history. The presence of a significant old population in the LMC is also consistent with the SFH derived from \citet{Harris2009}, where a significant portion of the LMC's stellar mass was formed during a short period.
\citet{Rezaeikh2014}, using also LPVs, found a 10\,Gyr old single star formation epoch, while in the LMC bar they identified a second distinct star formation episode starting around 3\,Gyr ago. Both events are captured in our study. We also examined whether our sample could be split spatially (e.g. bar versus outer disc), as suggested by the spatially-resolved SFHs of \citet{Ruiz-Lara2020} and by this distinct bar episode of \citet{Rezaeikh2014}; however, our classification-confirmed Mira sample is too small to support such a subdivision without individual age bins becoming dominated by only a handful of stars, and we defer this test to future work with a larger sample (Sect.~\ref{conclusions}).
Interestingly, we see thet for the intermediate-age population ($\rm \sim$3--6\,Gyr) the SFR is low with no obvious evidence of a star burst in line with \citet{Mazzi2021} and \citet{Ruiz-Lara2020}. One major feature of our reconstructed SFH is a recent starburst with a peak at $\sim$1--2\,Gyr coinciding very nicely with the discovery of a young population at $\rm \sim 2\,Gyr$ in \citet{Mazzi2021}. A comprehensive study based on data from the {\it Hubble} Space Telescope, whilst suggesting significant spatial variations, confirms a star formation activity at $\sim$2\,Gyr \citep{Cohen2024}.

To summarise, based on the period-age relation of fundamental pulsating AGB stars, we can reconstruct the SFH of the LMC consistently with other studies using global CMD-fitting techniques. The two period-age calibrations show broad consistency in their overall shape but differ in the relative contributions of the ancient and intermediate-age populations, reflecting the dominant systematic uncertainty in this analysis. This discrepancy, which exceeds the statistical uncertainties by a factor of 2--3, shows that the choice of period-age calibration is the dominant systematic uncertainty in the Mira-based SFH reconstruction.

\section{Chemical abundances of AGB stars in the LMC} \label{abundances}

Spectroscopic surveys of the Clouds beased on AGB stars remain scarce. \cite{olsen2011} targeted $\sim$6\,000 giants, including many AGB stars, revealing a counter-rotating population of metal-poor stars potentially accreted from the SMC. \cite{vandermarel2002} exploited $\sim$1\,000 spectra of C-rich stars to provide key constraints on the LMC’s kinematics, geometrical structure, total mass, and tidal radius.

The Apache Point Observatory Galactic Evolution Experiment (APOGEE; \citealp{majewski2017}) produced high-resolution near-infrared spectra for $~$4\,000 AGB stars across the Clouds \citep{Nidever2020}, but the abundance of multiple elements, their trends and  interpretation using a chemical evolution model have been obtained only for RGB stars \citep{Nidever2020,hasselquist2021}.

The APOGEE Stellar Parameter and Chemical Abundances Pipeline (ASPCAP) has been specifically designed to determine stellar parameters and chemical abundances from APOGEE observations. It uses two steps to analyze the data: First, the pipeline performs a global fit to the entire APOGEE spectrum to determine the key stellar atmospheric parameters. This involves simultaneously solving for effective temperature, surface gravity, microturbulence, overall metallicity ([M/H]), $\alpha$-elements ([$\alpha/$M]), carbon ([C/M]), and nitrogen ([N/M]). With the stellar parameters fixed from the first stage, ASPCAP then derives the individual chemical abundances. It refocuses the fit on specific small wavelength regions ("windows") in the spectrum that are dominated by the spectral lines of a particular element. These windows are chosen because they contain spectral features that are dominated by transitions of the element of interest. The pipeline applies a mask to analyze only these specific portions of the spectrum for the abundance determination of each element. The "weights" for these windows, which can be assigned based on the reliability of the line or level of contamination from other features, are provided in the APOGEE data release.

\subsection{Sample}
We selected our sample of stars based on their equatorial coordinates and radial velocities in the same way it was done for the Gaia DR3 light curves (see Sect.\ref{data}). In order to ensure to have high quality data, we excluded stars with STARFLAGS corresponding to \texttt{BAD\_PIXELS}, \texttt{PERSIST\_HIGH}, \texttt{VERY\_BRIGHT\_NEIGHBOR}, ASPCAP flags with \texttt{NO\_ASPCAP\_RESULTS}, \texttt{STAR\_BAD} and $\mathrm {SNR} > 20$. We refer the reader to the APOGEE DR17 documentation\footnote{https://www.sdss4.org/dr17/} for more information. Our final sample consists of 567 stars. We cross-matched our sample with the Gaia LPV catalogue from \citet{Lebzelter2023} where we found 296 sources in common. We considered only those as AGB stars and classified the rest as RGB stars. As mentioned in Sect.~\ref{sect:SFH}, the typical completeness of Gaia is 90\% for recovering O-rich AGB stars.

\begin{figure}[!htbp]
 \center{\includegraphics[width=0.49\textwidth]{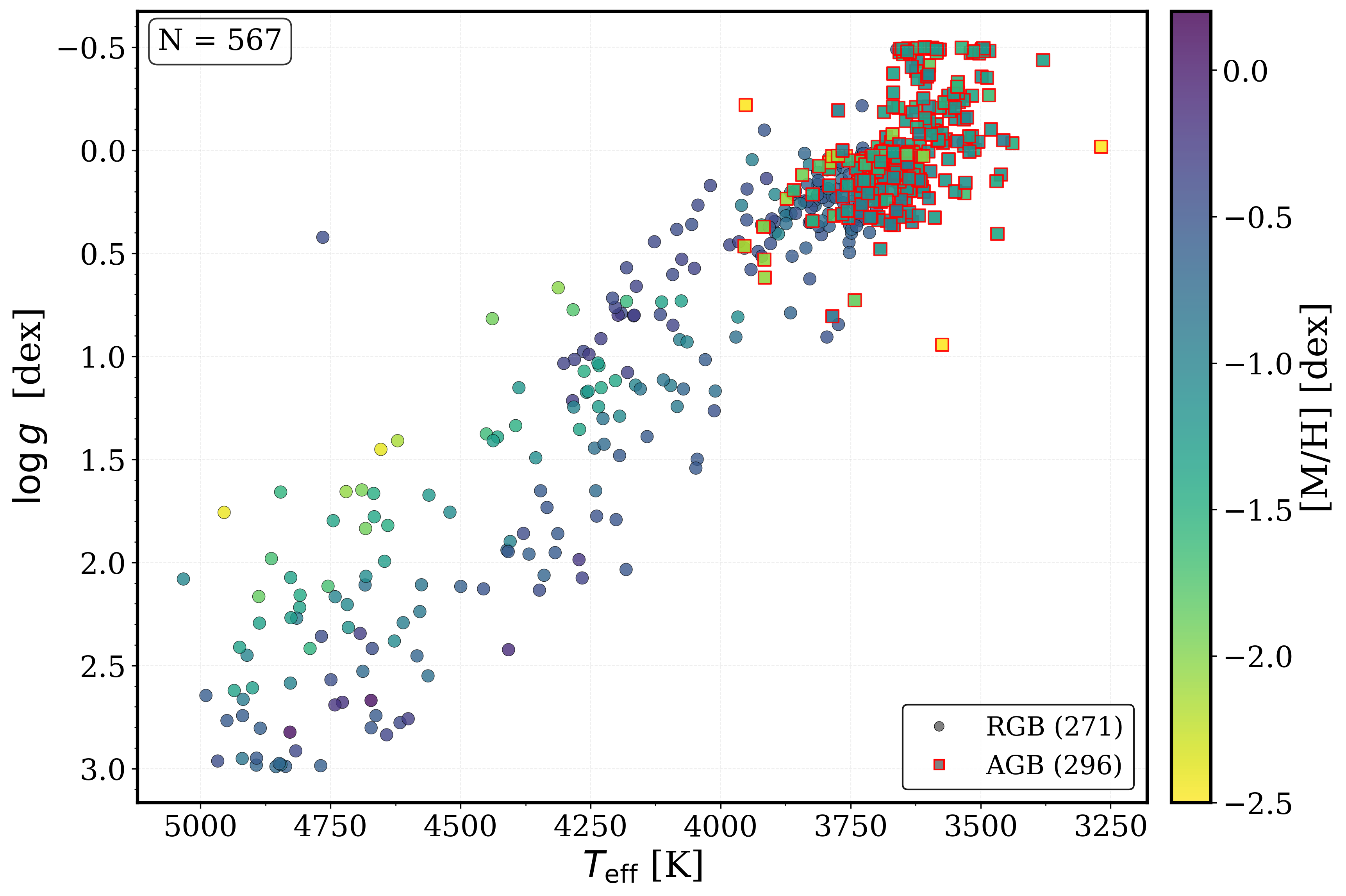} }
 \caption{Kiel diagram showing the distribution of RGB and AGB stars in the sample. Effective temperature ($\rm T_\mathrm{eff}$) is plotted against surface gravity (log\,g), with the color scale representing the stellar metallicity [M/H]. RGB stars are shown as circles (N = 271), while AGB stars are shown as squares (N = 296) with red edges. The color gradient from blue to yellow represents the metallicity range from $\rm [M/H] \sim  -2.5\, to +0.2$ dex.}

 \label{classification}
 \end{figure}

Figure~\ref{classification} shows the Kiel diagram  of our AGB/RGB star sample where the data reveal a clear separation between the two populations, with AGB stars generally occupying a region of lower effective temperatures ($\rm T_\mathrm{eff} \leq 4000 K$) compared to RGB stars. The mean metallicities of the RGB stars is $\rm [M/H] = -0.77 \pm 0.42$ dex while it is $\rm [M/H] = -0.79 \pm 0.22$ dex for the AGB stars showing no systematic difference between these two populations.

\subsection{Chemical abundances}

As in APOGEE we run the FERRE code with an improved version where we perform line-to-line individual abundances. We use the APOGEE input spectra together with their derived  stellar parameters to feed in our pipeline. The algorithm begins by scanning through all wavelength positions in the mask file and identifying pixels where the target element's weight meets the minimum threshold criterion. We have chosen the minimum weight to be 0.01 allowing to exploit weak lines. Secondly, the merging algorithm examines adjacent clusters and determines whether they should be merged based on the separation between their central wavelengths. Based on the instrumental profile of APOGEE, the threshold is set at $\rm 2.10\,\AA$. Then we apply a quality filter where we calculate the line-to-line dispersion using a threshold of 0.2\,dex. Lines with insufficient measurements or excessive scatter are rejected. Furthermore, measurements falling exactly on the boundaries of the synthetic grid (--0.5 dex and +0.75 dex, respectively) are also excluded. Since these edge values represent failed fits rather than true physical measurements, their removal is necessary not to affect the line-to-line dispersion computations.
We specify in Tab.~\ref{tab:lines} the mean abundance for each line together with its standard  deviation. Overall, we do not notice any significant differences in the mean abundance between RGB and AGB stars in our sample. 

\begin{figure*}[!htbp]
 \center{\includegraphics[width=0.98\textwidth]{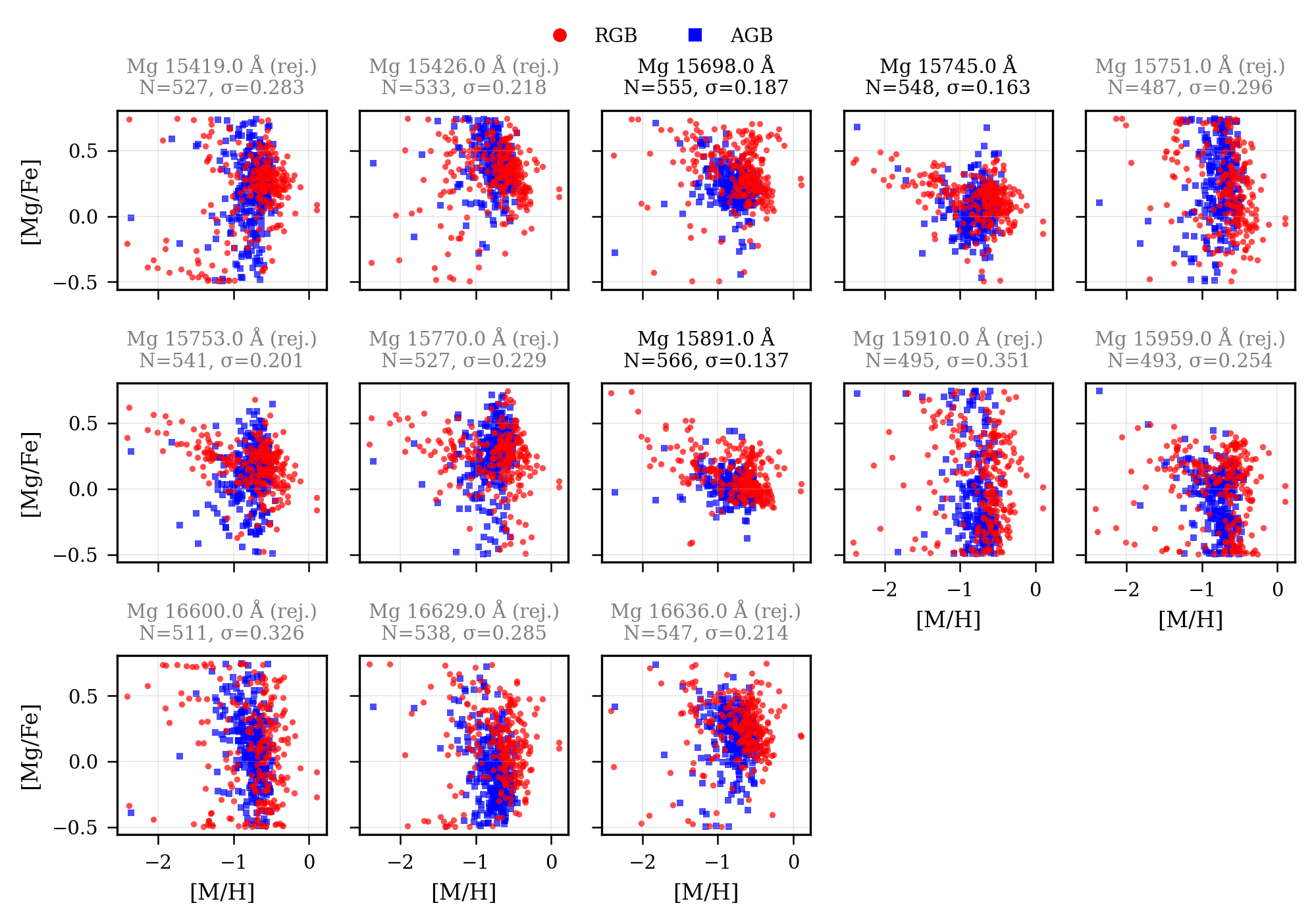} }
 \caption{Individual Mg lines as a function of [M/H] for our sample of RGB stars (red) and AGB stars (blue). For each line the standard deviation $\sigma$ is indicated, along with the number of stars N for which a reliable measurement was found. Rejected lines are marked with "rej.".} Only the Mg line at 15\,698\,\AA\,, 15\,745\,\AA\, and 15\,891\,\AA\, are used in our analysis while the others are discarded.
 \label{MgFe}
 \end{figure*}

\begin{figure*}[!htbp]
\center{\includegraphics[width=0.99\textwidth]{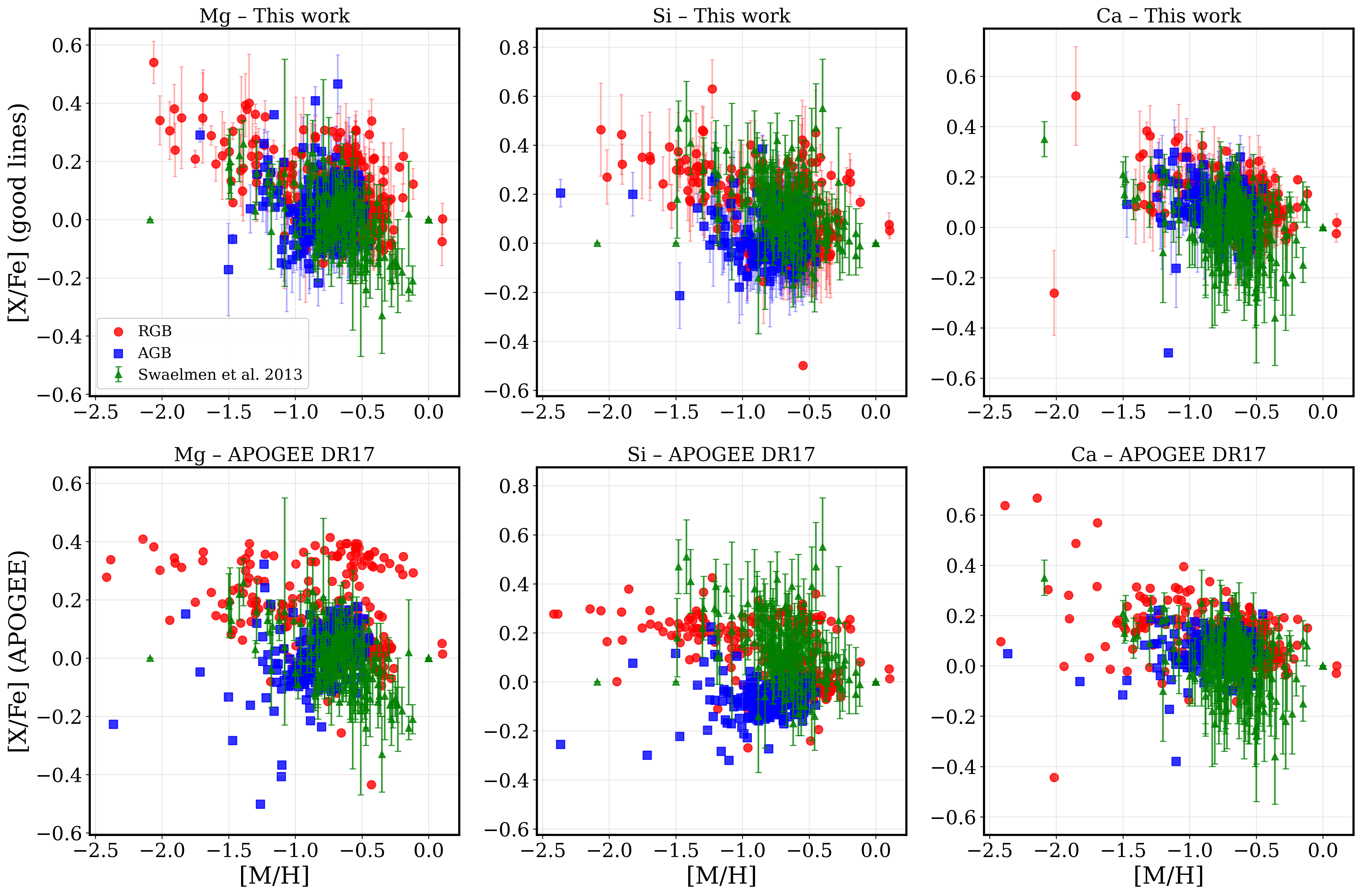}}
 \caption{Comparison of our  derived Mg, Si and Ca abundances (upper left, middle, right panels, respectively) from selected lines vs the total Mg, Si and Ca abundances from APOGEE DR17 (lower panels). Red stars are RGB stars, blue ones are AGB stars. The error bars show the line-to-line uncertainties from our selected lines. As a comparison sample the abundances derived from optical FLAMES observations of RGB stars  (\citealt{swaelmen2013}) are overplotted.} 
 \label{combined_abundances}
 \end{figure*}

Figure \ref{MgFe} shows each individual Mg line plotted against [M/H].
The majority of the lines show a large intrinsic dispersion; hence those lines can not be trusted. Some lines, e.g. the Mg line at 15\,419\,\AA\, or at 16\,600\,\AA \, show just a large dispersion  with no trend of [Mg/Fe] with metallicity.
This could be due mainly to blends with strong molecules (e.g. CN, CO and OH) for these cool objects. Finally, for Mg only three lines (at 15\,698\,\AA\,, 15\,745\,\AA\, and 15\,891\,\AA\,) fulfill our criteria and are used for our analysis (see Tab.~\ref{tab:lines}).
Figure~\ref{combined_abundances} compares our Mg abundances, derived from selected lines, with the total Mg abundances from the optical sample of \citet[left panel]{swaelmen2013} and from APOGEE DR17 (lower left panel): the smaller dispersion in the Mg abundances at a given metallicity and the good agreement with the optical sample are clearly visible. Our line-by-line Mg abundances show a scatter about 25\% smaller than that of the APOGEE DR17 pipeline values.

In addition, we do not notice any trends of the Mg abundances related to the effective temperature (Fig. \ref{combined_feh_teff}, left panels), indicating that AGB stars can be analysed in this way without introducing biases in the Mg abundance.

We refer to Fig. \ref{SiFe} and Tab. \ref{tab:lines} for details about the adopted Si lines. Compared with the total Si abundances from APOGEE DR17 (Fig. \ref{combined_abundances}, central lower panel), our abundances show a dispersion about 10\% smaller than the APOGEE DR17 values. We find a trend of the Si abundances in the sense that cooler temperatures reveal lower Si abundances. Also, the temperature effect in APOGEE DR17 seems to be slightly larger than in our analysis (see Fig.~\ref{combined_feh_teff}, middle panels).

\begin{table}[!htbp]
    \centering
    \begin{tabular}{c|cccc}
    element & $\lambda$ ($\AA$) & $\sigma$ (dex) & RGB & AGB\\
    \hline
    Mg     & 15698 & 0.187  & 0.114 $\pm$ 0.12 & 0.11 $\pm$  0.09\\
    Mg     & 15745 & 0.163  & 0.096 $\pm$ 0.09 & 0.039 $\pm$  0.18\\
    Mg     & 15891 & 0.137  & 0.101 $\pm$  0.17 & 0.018 $\pm$  0.09\\
    Si     & 15889 & 0.173  & 0.121 $\pm$  0.22 & 0.058 $\pm$  0.16\\
    Si    & 15893 & 0.141  & 0.225 $\pm$  0.15 & 0.114 $\pm$  0.13\\ 
    Si    & 16247 & 0.167  & 0.027 $\pm$  0.19 & 0.104 $\pm$  0.12 \\
    Si    & 16385 & 0.191  & 0.106 $\pm$ 0.18 & 0.04 $\pm$  0.14\\
    Ca   & 16141 & 0.159  & 0.041 $\pm$  0.15 & 0.043 $\pm$  0.15 \\
    Ca   & 16155 & 0.137 & 0.082 $\pm$  0.14 & 0.077 $\pm$  0.15\\
    Ca & 16161 & 0.117 & 0.142 $\pm$  0.10 & 0.140 $\pm$  0.11 \\
    \hline
    \end{tabular}
    \caption{Individual lines used for our analysis: the second column gives the central wavelength and the third column the typical standard deviation based on the number of stars for which a reliable measurement was found. The fourth and fifth columns specify the mean abundance and its standard deviation, respectively.}
    \label{tab:lines}
\end{table}

We were able to successfully exploit all three Ca lines within our observed spectra, since they provided good fits (see Fig. \ref{CaFe} and Tab. \ref{tab:lines} for more information). We see here similar patterns as for the other $\alpha$-elements: an improvement in the intrinsic dispersion of the Ca abundances compared to APOGEE DR17 (Fig.~\ref{combined_abundances}, lower right panel), with a scatter about 10\% smaller, and no signs of a temperature trend (Fig.~\ref{combined_feh_teff}, right panels).

The individual-line Mg, Si, and Ca abundances for all 567 stars in our sample, together with the adopted [Mg/Fe], [Si/Fe], and [Ca/Fe] values used throughout this paper, are made available in their entirety as a machine-readable table at the CDS (Table~\ref{tab:cds}). Table~\ref{tab:cds} is published in its entirety at the CDS via anonymous ftp to cdsarc.cds.unistra.fr (130.79.128.5) or via \url{http://cdsarc.u-strasbg.fr/cgi-bin/qcat?J/A+A/}; a portion is shown here for guidance regarding its form and content.

\begin{table*}[!htbp]
    \centering
    \small
    \begin{tabular}{lcccccccccccc}
    \hline\hline
    ID & Teff (K) & log\,g & [M/H] & Type & Mg15698 & [Mg/Fe] & Si16247 & [Si/Fe] & Ca16141 & [Ca/Fe] & ... \\
    \hline
    2M05434720-6644121 & 4081.68 & 2.034 & $-$0.523 & RGB & 0.516 & 0.041 & 0.231 & 0.249 & 0.097 & 0.067 & ... \\
    2M04413697-6818530 & 3555.29 & 0.009 & $-$0.873 & AGB & 0.282 & 0.043 & $-$0.230 & $-$0.167 & 0.020 & 0.111 & ... \\
    2M04324400-6937420 & 3536.81 & 0.199 & $-$0.961 & AGB & 0.163 & 0.080 & $-$0.063 & $-$0.038 & 0.023 & 0.088 & ... \\
    ... & ... & ... & ... & ... & ... & ... & ... & ... & ... & ... & ... \\
    \hline
    \end{tabular}
    \caption{Line-by-line Mg, Si, and Ca abundances. One representative individual-line column is shown per element (Mg15698, Si16247, Ca16141) together with the adopted [Mg/Fe], [Si/Fe], and [Ca/Fe] values; the full table, including the abundance derived from every one of the 13 Mg, 26 Si, and 3 Ca lines considered, is available in machine-readable form at the CDS.}
    \label{tab:cds}
\end{table*}

\subsection{Chemical abundances of AGB for tracing the intermediate-age populations}

As AGB stars are large-amplitude pulsators and the stellar atmospheres are dominated by dynamical processes, we investigate here for the three $\alpha$-elements if we see any correlation with stellar parameters, pulsation period and amplitude of the star. 

\begin{figure}[!htbp]
 \center{\includegraphics[width=0.49\textwidth]{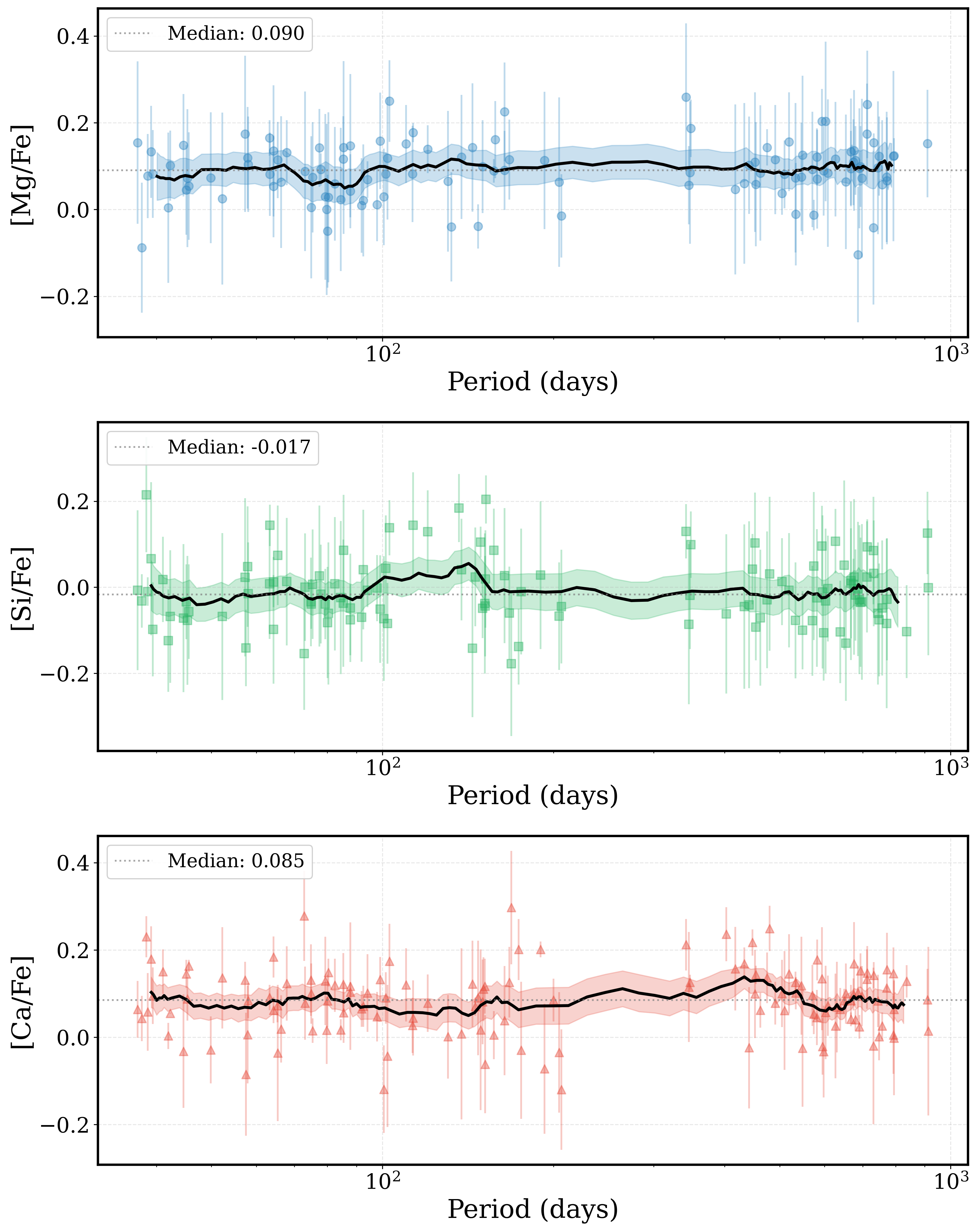} }
 \caption{Abundance trends as a function of pulsation period for Mg (top), Si (middle), and Ca (bottom). Individual measurements are shown as colored points with error bars representing the line-to-line dispersion (standard deviation among multiple spectral lines for each element). The black solid line shows the running mean within a window of 15 points, and the shaded region indicates the combined $\rm \pm 1\,\sigma$ uncertainty. Horizontal dashed lines mark the median abundance for each element.} 
 
 \label{periodvsabundance}
 \end{figure}

To study whether the chemical abundances of Mg, Si, and Ca depend on stellar pulsation period, we examined the relationship between these key $\alpha$-elements and the fundamental pulsation period derived from our data set. Figure \ref{periodvsabundance} presents the abundance distributions as a function of period for Mg (top panel), Si (middle panel), and Ca (bottom panel). A running mean analysis reveals a relatively flat trend across the entire period range for all elements with no sign of a correlation between the period and the abundance. The mean abundances  of Mg are $\rm 0.020\, \pm 0.103$ dex, of Si  $\rm 0.015\, \pm 0.093$ dex, and  of Ca  $\rm 0.073\, \pm 0.088$ dex showing similar trends within the uncertainties. We also do not see any correlation between the amplitude of the star and the abundances nor with the stellar parameter such as $\rm T_{eff}$, $\rm log\,g$, and $\rm [M/H]$. We therefore conclude that AGB stars (O-rich Miras) can be used to trace chemical abundances and therefore the SFH. However, a careful selection of lines suitable for the analysis is essential.

\citet{Nidever2020} and \citet{hasselquist2021} produced chemical evolution models to fit the chemical abundances of RGB stars derived from APOGEE spectra and several elements. They found a SFH peak 2--3\,Gyr ago in agreement with a previous chemical evolution study by  \citet{bekki2012} who only looked at Ba and Mg abundances, also for RGB stars. The latest chemical evolution models use as input the  SFH from external data (e.g. from CMD fttting). For example, \citet{Vasini2023} found that the SFH by \citet{Harris2009} best reproduced the current SFR, supernova and nova rates, gas mass and metallicity of the LMC from their chemical evolution models. More recently,  \citet{Palla2026} fitted r- and s-process elements in the SMC  adopting the SFH from the VMC survey  (\citealt{rubele2018}). In a future work we will compute specifically trimmed chemical evolution models including our derived SFH (Sect.~\ref{sect:SFH}) together with our improved analysis of chemical abundances of the intermediate-age population.

\section{Conclusions and future perspectives} \label{conclusions}

In this paper, we have presented a first study of the SFH and chemical abundances of the LMC, with a particular emphasis on the AGB stellar population—an intermediate-age tracer that has been largely overlooked in previous stellar population studies. This is the first work from the Clouds in the "Middle Ages" project that uses AGB stars to find and characterise kinematical and chemical patterns associated to the SFH of the Clouds.

We reconstructed the SFH of the LMC using a chemistry-inclusive sample of 2047 fundamental-mode pulsating AGB stars (879 O-rich, 1168 C-rich) identified from the Gaia DR3 LPV catalog, applying two independent period-age calibrations: the theoretical relation of \citet{Trabucchi2022} and the empirical calibration of \citet{Zhang2023}. Our analysis reveals significant systematic differences between these calibrations, particularly at ages >3 Gyr where the \citet{Trabucchi2022} relation shows some star formation only $3-6$ Gyr ago while the \citet{Zhang2023} relation predicts a prominent event >6 Gyr ago. Neither calibration reproduces the independent CMD-based SFH determinations (\citealt{Ruiz-Lara2020}, \citealt{Mazzi2021}) equally well across all ages, so we do not adopt either relation as a single fiducial calibration; instead we present the SFH under both (Figs.~\ref{fig:sfh_all} and~\ref{fig:sfh_mira}) and treat their difference as our primary estimate of the systematic uncertainty. An old-age population, and the disagreement between calibrations more generally, is recovered consistently whether we use our amplitude-selected Mira-only sample or the full fundamental-mode-pulsator sample. We identify three distinct phases of star formation: an old population at $\rm \sim$ 7--10 Gyr contributing to $\rm \sim 20\%$ of the total stellar mass formed, a quiescent intermediate-age phase at 3--6 Gyr with no evidence of major bursts, and a recent starburst at $\rm \sim$ 1--2 Gyr. These findings demonstrate that Mira variables can serve as effective tracers of the LMC's SFH when carefully calibrated, though the choice of period-age relation remains the dominant systematic uncertainty.

Our chemical abundance analysis of 567 RGB and AGB stars observed by APOGEE, using an improved line-by-line abundance pipeline, demonstrates that AGB stars can reliably trace chemical enrichment despite their pulsational variability and complex atmospheric dynamics. We identified high-quality spectral lines for Mg (15\,698\,\AA\,, 15\,745\,\AA\,, 15\,891\,\AA\,), Si (15\,889\,\AA\, 15\,893\,\AA\,, 16\,247\,\AA\,, 16\,385\,\AA\,), and Ca (16\,141\,\AA, 16\,155\,\AA, 16\,161\,\AA) that yield significantly reduced intrinsic scatter compared to the APOGEE DR17 pipeline results. Crucially, we find no statistically significant correlations between the derived $\alpha$-element abundances ([Mg/Fe], [Si/Fe], [Ca/Fe]) and pulsation period, pulsation amplitude, or stellar parameters such as $\rm T_{eff}$, log\,g, or $\rm [M/H]$. The mean abundances for the AGB sample ($\rm [Mg/Fe] = 0.020 \pm 0.103$ dex, $\rm [Si/Fe] = 0.015 \pm 0.093$ dex, $\rm [Ca/Fe] = 0.073 \pm 0.088$ dex) are consistent with those of RGB stars and with the optical sample of \citet{swaelmen2013}, indicating that AGB stars preserve the chemical signatures of their progenitor populations without substantial bias introduced by the pulsation phenomenon.

Looking ahead, the 4-metre Multi-Object Spectroscopic Telescope facility (4MOST; \citealp{dejong2019}), mounted at the 4m-class telescope VISTA, will allow us to make a significant step forward in using AGB stars to probe the formation and evolution of the Clouds. In particular, the 1001 Magellanic fields survey (1001MC; \citealp{cioni2019}), a survey of the 4MOST consortium, will provide high-resolution, multi-object spectroscopy for thousands of stars across the entire LMC and SMC. This will enable detailed abundance analysis of AGB stars across all spatial regions, including the disc, bar, and outer halo, allowing us to map the chemical signatures of past interaction events with unprecedented precision. Combined with precise Gaia DR4 proper motions, and an enlarged sample of LPVs, these data will also permit dynamical age estimates independent of period-age relations, offering a powerful cross-calibration of the Mira period-age relation and a robust constraint on the LMC-SMC-Milky Way interaction history.
We expect the number of LPVs in {\it Gaia} DR4 to increase by a factor of three with a major improvement on the number of small amplitude variables beyond the DR3 lower limit of  the amplitude in $\rm G \ge 0.1$\,mag.

\begin{acknowledgements}
This work was supported by the Deutsche Forschungsgemeinschaft (DFG, German Research Foundation) -- [Project number 550131653]. This work is in the honor of Prof. Paola Marigo, a leader in the study of AGB stars and esteemed colleague, who left us prematurely.
\end{acknowledgements}

\bibliography{lmc.bib}

\begin{appendix}

\section{Amplitude classification} \label{amplitude}
\begin{figure}
    \centering
    \includegraphics[width=0.49\textwidth]{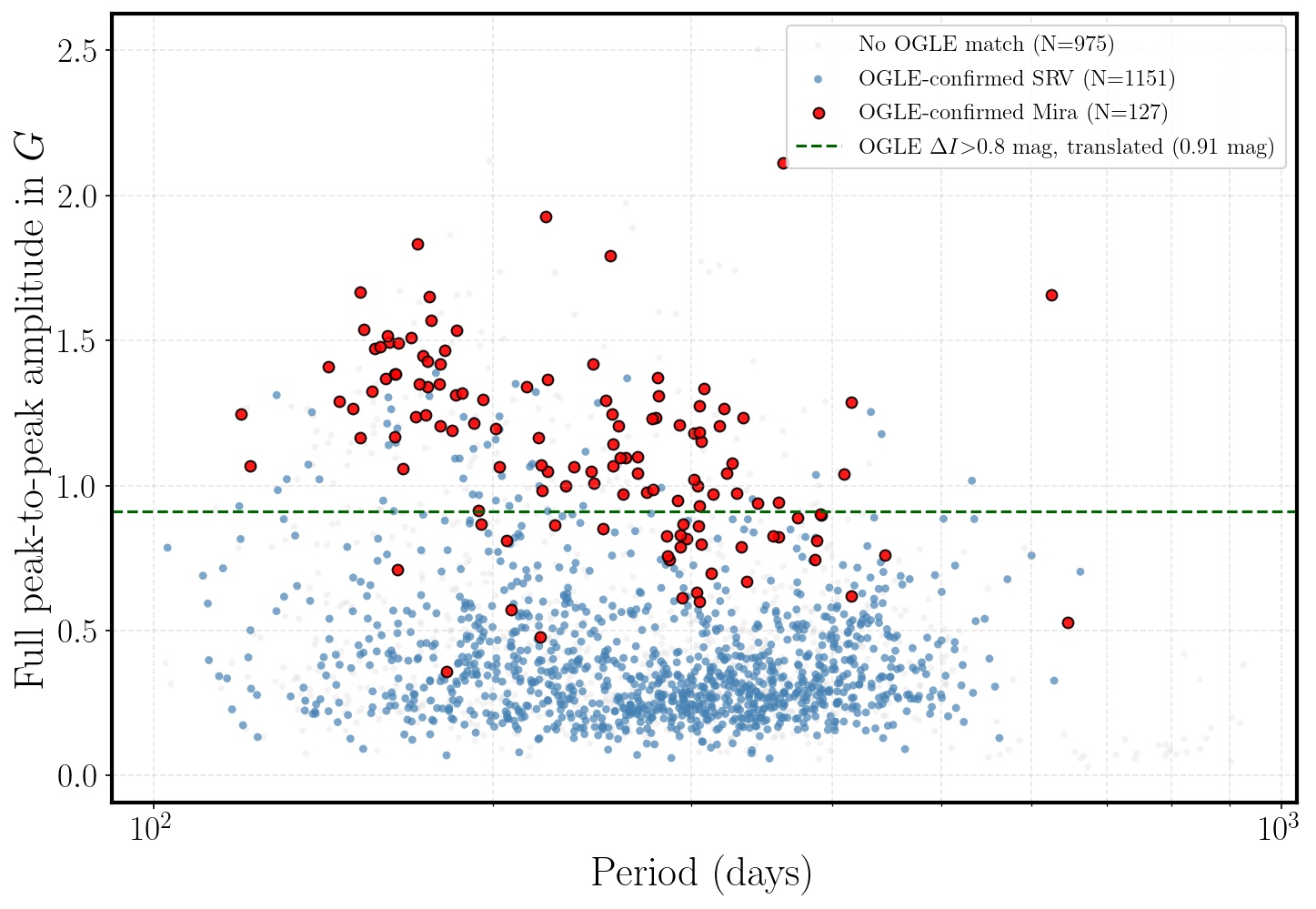}
    \caption{Period vs. full peak-to-peak $G$ amplitude for the fundamental-mode sample, colour-coded by OGLE classification (confirmed Mira / confirmed SRV / no OGLE match), with the translated $\Delta G \gtrsim 0.91$ mag boundary marked}
    \label{fig:period-amplitude}
\end{figure}

\begin{figure}
    \centering
    \includegraphics[width=0.49\textwidth]{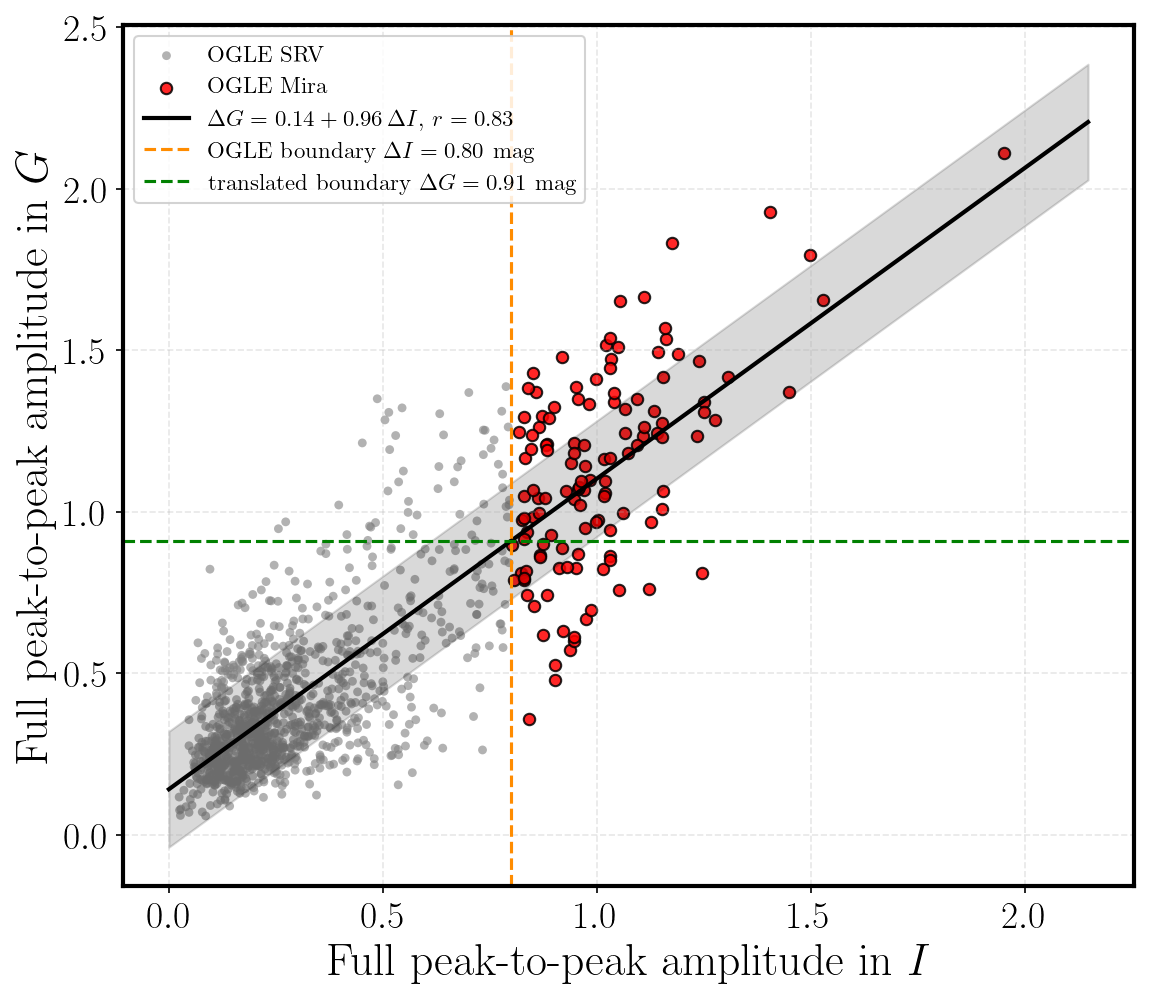}
    \caption{The fitted $\Delta G$ vs. $\Delta I$ relation itself, with the OGLE $\Delta I = 0.8$ mag boundary and its translated $\Delta G$ counterpart marked, points colour-coded by OGLE Mira/SRV classification}
    \label{fig:amplitude-gi}
\end{figure}

We show the selection of Mira variables from the amplitude in the $G$ band as translated from the amplitude defined in the OGLE-III study \citep{soszynski2009}. Figure \ref{fig:period-amplitude} shows the $G$-band amplitude vs. period while Fig.\,\ref{fig:amplitude-gi} vs. OGLE-$I$-band amplitude.

\section{Si abundances}
We show the line-to-line comparison of Si abundances. 

\begin{figure*}[!htbp]
 \center{\includegraphics[width=0.98\textwidth]{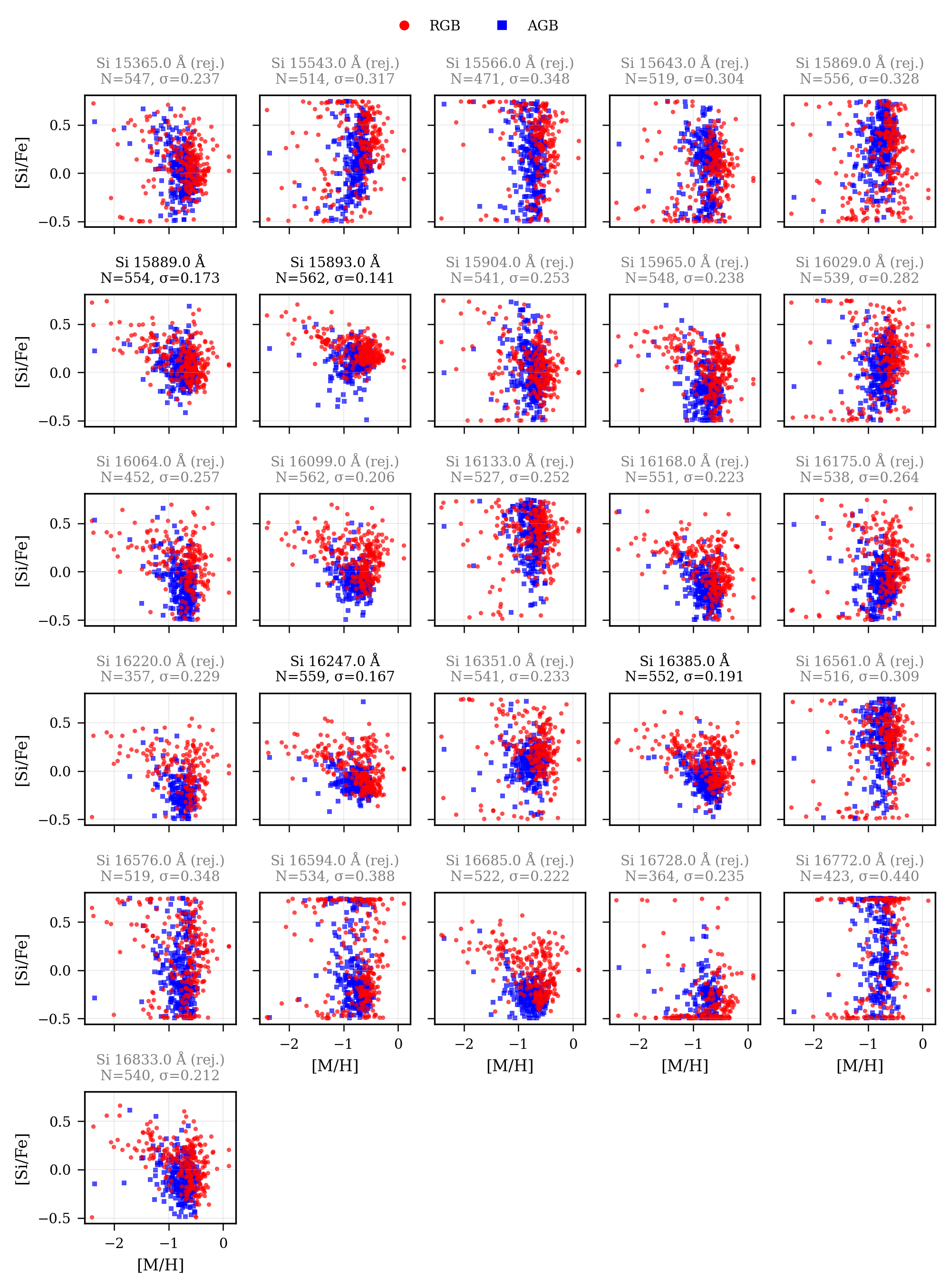} }
 \caption{As Fig.\,\ref{MgFe} but for Si lines.}
 \label{SiFe}
 \end{figure*}
 
\section{Ca abundances}
We show the line-to-line comparison of Ca abundances. 

\begin{figure*}[!htbp]
 \center{\includegraphics[width=0.98\textwidth]{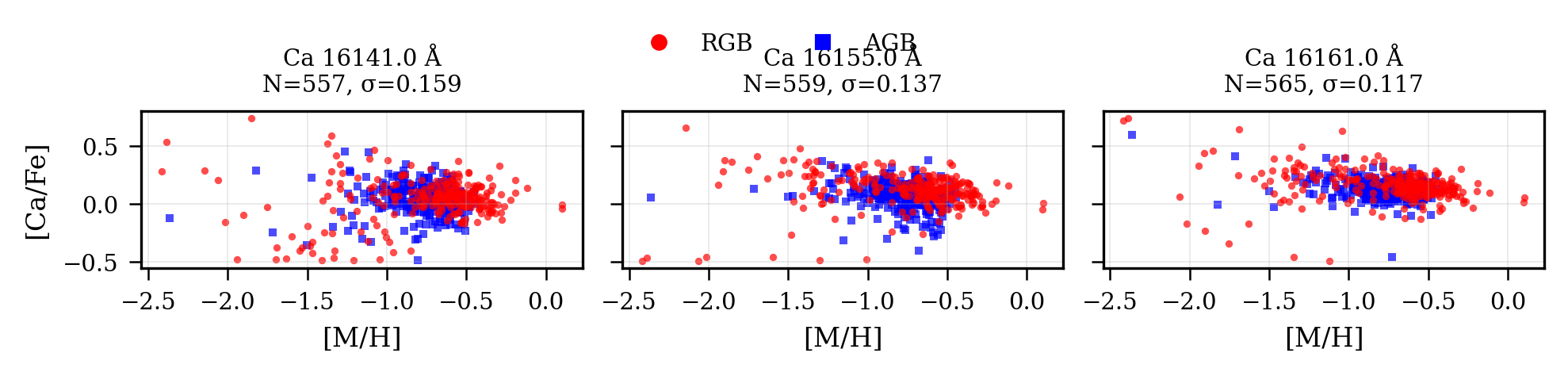} }
 \caption{As Fig.\,\ref{MgFe} but for Ca lines.}
 \label{CaFe}
 \end{figure*}

\section{Temperature dependence of $\alpha$-elements}
 We show the temperature dependence of the elements [Mg/Fe], [Si/Fe] and [Ca/Fe] vs. [Fe/H] for the AGB and RGB stars in our sample.
 
\begin{figure*}[!htbp]
 \center{\includegraphics[width=0.99\textwidth]{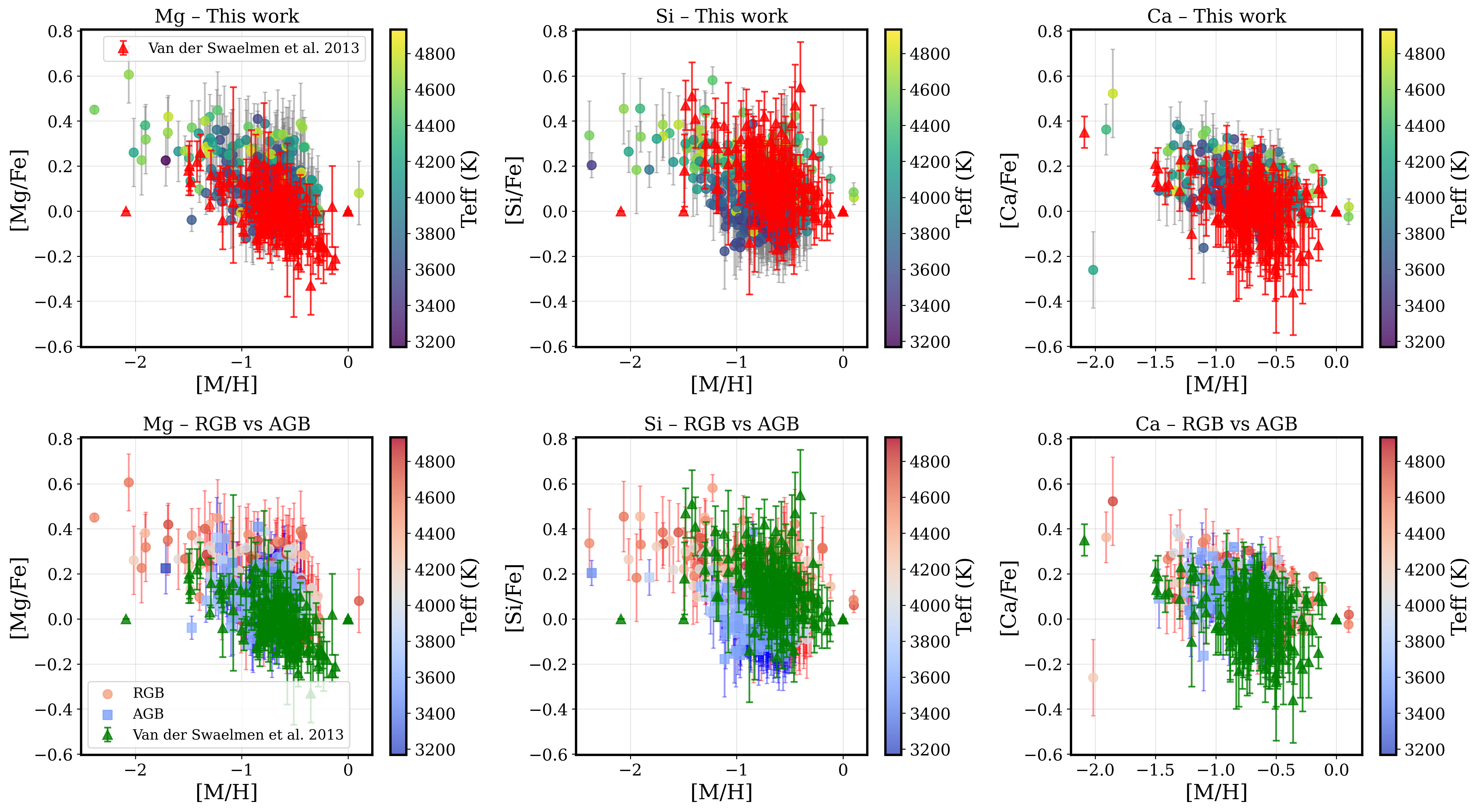} }
 \caption{(Upper panel): [X/Fe] vs. [Fe/H] as a function of effective temperature, with [Mg/Fe] on the left, [Si/Fe] in the middle, and [Ca/Fe] on the right. The error bars give the line-to-line dispersion of our analysis. In red the comparison sample of \citet{swaelmen2013}. (Lower panel): Same as upper panel but for our RGB (filled circles) and AGB (filled squares) samples, separately}
 \label{combined_feh_teff}
 \end{figure*}

 \end{appendix}
\end{document}